\documentclass[preprint]{elsarticle}
\usepackage[utf8]{inputenc}
\usepackage{amsmath}
\usepackage{graphicx}
\usepackage{amssymb}
\usepackage{subfigure}
\usepackage{caption}
\usepackage{float}
\usepackage{siunitx}
\usepackage{algpseudocode} 
\usepackage{multirow}
\usepackage[toc]{appendix}
\usepackage{array}
\usepackage{rotating} 
\usepackage{tikz}
\usepackage{ctable}
\usepackage{hyperref}
\begin{document}

\title{What is in the model? A Comparison of variable selection criteria and model search approaches}

\author[1]{Shuangshuang Xu}
\affiliation[1]{organization={Fralin Biomedical Research Institute},
city={Roanoke},
addressline={Virginia},
postcode={24016},
country={U.S.A.}}

\author[2]{Marco A. R. Ferreira}
\affiliation[2]{organization={Department of Statistics, Virginia Tech},
city={Blacksburg},
addressline={Virginia},
postcode={24060},
country={U.S.A.}}

\author[1]{Allison N. Tegge \corref{cor1}}
\ead{ategge@vt.edu}
\cortext[cor1]{Corresponding author}

\begin{abstract}
For many scientific questions, understanding the underlying mechanism is the goal. 
To help investigators better understand the underlying mechanism, variable selection is a crucial step that 
permits the identification of the most associated regression variables  of interest.
A variable selection method consists of model evaluation using an information criterion and a search of the model space.
Here, we provide a comprehensive comparison of variable selection methods 
using performance measures of  correct identification rate (CIR), recall, and false discovery rate (FDR).
We consider the
BIC and AIC for evaluating models, and exhaustive, greedy, LASSO path, and
stochastic search approaches for searching the model space; we also consider LASSO using
cross validation.
We perform simulation studies for linear and generalized linear models that parametrically explore a wide range of realistic sample sizes, effect sizes, and correlations among regression variables. We consider  model spaces with a small and larger number of potential regressors. 
The results show that the exhaustive search BIC and stochastic search BIC outperform the other methods when considering the performance measures on small and large model spaces, respectively. These approaches result in the highest CIR and lowest FDR, which collectively may support long-term efforts towards increasing replicability in research.
\end{abstract}


\maketitle

\section{Introduction}

Screening and identifying the regression variables that are most relevant  to a research question is essential~\citep{guyon2003introduction}. The importance of this notion has been amplified in the real-world data landscape where the expansion of storage capabilities and advancements in data assessment techniques have tremendously increased the collection of an abundance of measurements and variables~\citep{fan2014challenges}. 
However, not all these variables are critical in answering specific research questions, and thus variable selection techniques need to be utilized~\citep{heinze2018variable}. 
Here, we study the interplay between variable selection criteria and methods to explore the model space while considering a wide array of sample sizes and other influential parameter settings. 

Establishing final models that only include truly important variables has at least three advantages. 
First, the elimination of superfluous variables can increase the robustness of conclusions across different samples, thus increasing replicability~\citep{meinshausen2010stability}. 
Second, the identification of a parsimonious model enhances interpretability of the findings. That is, a simpler model with fewer variables is easier to understand, which helps deciphering the underlying processes that govern the observed phenomena~\citep{hastie2015statistical}. 
Third, the inclusion of extraneous variables in the model can affect the estimation of the true parameters of the model \citep{hurtz2000personality}. For example, the inclusion of extraneous variables that are correlated with true variables will result in inflated variance of coefficient estimates of the true variables.
 
One way to identify the important variables is through variable selection. Variable selection often uses a two-pronged approach that relies on searching the model space and evaluating putative models through an information criterion.
The search through the model space is a crucial part of variable selection. 
An exhaustive search of the model space is the most comprehensive, although  time consuming and only feasible when a small number of predictors are considered~\citep{nievergelt2000exhaustive}. For larger model spaces, a greedy search, such as stepwise selection, is often utilized~\citep{zhang2016variable}. Unfortunately, greedy approaches may get stuck at a local optimal model rather than the global optimal model~\citep{simmons2019beware}. To overcome this limitation, stochastic search \citep{golberg1989genetic, kirkpatrick1983optimization,scrucca2013ga} options have been used to deal with high-dimensional optimization. 

The second part of variable selection is evaluating the quality of a putative model. For this, an information criterion is typically used, with Akaike Information Criterion (AIC; \cite{akaike1974new,akaike1998information}) and the Bayesian Information Criterion (BIC; \cite{schwarz1978estimating}) as two of the most popular approaches. These information criteria seek to balance the trade-off between how complex the model is with how well the model fits the data in an effort for selecting a parsimonious model. 
There are several other information criteria, such as Deviance information criterion (DIC, \cite{spiegelhalter2002bayesian}) and Watanabe-Akaike information criterion (WAIC, \cite{watanabe2010asymptotic}); DIC and WAIC are usually used for comparison of complex models that include random effects.

Another group of approaches for variable selection include likelihood penalization based methods.
The most widely used of these methods, the Least Absolute Shrinkage and Selection Operator (LASSO; \cite{tibshirani1996regression}), 
performs variable selection by incorporating an $L_1$ penalty in the objective function to reduce the coefficients of less important variables to  zero. 
In particular, LASSO requires a regularization parameter that multiplies the $L_1$ penalty; 
selecting this parameter is non-trivial and may greatly impact LASSO's performance.
Implementation of LASSO usually considers a grid of regularization parameter values, each leading to a selected model (i.e., the LASSO path)~\citep{hastie2009elements}.
Thus, when considering the set of models in the LASSO path, LASSO can be seen as a method for the exploration of the model space. 
Additional regularization approaches include ridge regression \citep{hoerl1970ridge} and elastic net \citep{zou2005regularization}. 
These approaches are very efficient for the exploration of large model spaces.

Despite the abundance of variable selection methods for different settings, considerations about the interplay between the different model search approaches and model evaluation criteria are lacking. In this paper, we seek to provide recommendations for when to use which search and evaluation criteria
through a comprehensive comparison of various variable selection approaches and  simulation studies.
We focus on variable selection for linear models (LM) and generalized linear models (GLM). We explore  exhaustive, greedy, LASSO path, and stochastic search of the model space. 
We compare model evaluation criteria BIC and AIC. Additionally, we include a comparison with the traditional implementation of LASSO with cross validation to evaluate its performance against these methods in both small- and large-dimensional settings.

The remainder of the paper is organized as follows. Section~\ref{sec:2} introduces the models and notation we consider. Section~\ref{sec:3} presents the variable selection methods. Section~\ref{sec:4} presents the results of simulation studies. Section~\ref{sec:package} describes the R package we have developed to support this work. Section~\ref{sec:5} presents the discussion and conclusions.

\section{Models} \label{sec:2}
We focus on variable selection for LMs and GLMs. LMs assume that the dependent variable is normally distributed and the relationship between the dependent variable and the regressors is linear. GLMs extend the traditional LMs to allow for more flexible distributions for the dependent variable, such as binomial, Poisson, and exponential distributions. Hence, GLMs can be applied to more data types.

\subsection{Linear model}

An LM can be expressed as:
\begin{equation} \label{eq:linear-model}
    \mathbf{y} = \mathbf{X} \boldsymbol{\beta} + \boldsymbol{\epsilon},
    \end{equation}
where $\mathbf{y}$ is an $n \times 1$ vector of the observed dependent variable for a sample of $n$ observations, $\mathbf{X}$ is an $n \times p$ matrix of $p$ regression variables, $\boldsymbol{\beta}$ is a $p \times 1$ vector of regression coefficients, and $\boldsymbol{\epsilon}$ is an $n \times 1$ vector of errors. Assuming the errors are independent and normally distributed with mean 0 and constant variance $\sigma^2$, the likelihood function is
\[
p(\mathbf{y} | \mathbf{X}, \boldsymbol{\beta}, \sigma^2) = \prod_{i=1}^{n} \frac{1}{\sqrt{2 \pi \sigma^2}} \exp \left( -\frac{(y_i - \mathbf{x}_i^\top \boldsymbol{\beta})^2}{2 \sigma^2} \right).
\]
Hence, the log-likelihood function is
\begin{equation} \label{eq:log-like-lm}
     \ell (\boldsymbol{\beta}, \sigma^2 | \mathbf{y}) = -\frac{n}{2} \log(2 \pi) - \frac{n}{2} \log(\sigma^2) - \frac{1}{2 \sigma^2} (\mathbf{y} - \mathbf{X} \boldsymbol{\beta})^\top (\mathbf{y} - \mathbf{X} \boldsymbol{\beta}),
\end{equation}
which is used both to find the estimates of the parameters and to quantify model fit.

\subsection{Generalized linear model}

A GLM assumes that the response variable $y_i,i=1\dots n$ follows a distribution from the exponential family with probability density function
\[
p(y_i | \theta_i, \phi) = \exp\left( \frac{y_i \theta_i - b(\theta_i)}{a(\phi)} + c(y_i, \phi) \right),
\]
where $\theta_i$ is the canonical parameter and $\phi$ is the dispersion parameter. In addition, $a(\cdot)$, $b(\cdot)$, and $c(\cdot)$ are known functions that define the distribution. 
Lastly, a GLM assumes that the expected value of $y_i$ is connected to the vector of regression variable $\mathbf{x}_i$ through the link equation 
\begin{equation} \label{eq:glm-model}
g(\mathbb{E}[y_i | \mathbf{x}_i]) = \mathbf{x}_i^\top \boldsymbol{\beta},
    \end{equation}
where $g(\cdot)$ is a link function that relates the expected value of $y_i$ to the linear predictor $\mathbf{x}_i^\top \boldsymbol{\beta}$.  Hence, the log-likelihood function is
\begin{equation}\label{eq:log-like-glm}
\ell(\boldsymbol{\beta} | \mathbf{y}, \mathbf{X}) = \sum_{i=1}^{n} \left[ \frac{y_i \theta_i - b(\theta_i)}{a(\phi)} + c(y_i, \phi) \right].
    \end{equation}

Consider for example the logistic regression, also known as Bernoulli GLM, which is widely used for datasets with binary dependent variables. 
Specifically, logistic regression assumes \( y_i \sim \text{Bernoulli}( p_i) \) and a  logit link function such that \( \log\left({p_i}/(1 - p_i)\right) = \mathbf{x}_i^T \boldsymbol{\beta} \). In this case, the log-likelihood function is \begin{equation}
    \ell(\boldsymbol{\beta}) = \sum_{i=1}^n \left[ y_i \log(p_i) + (1 - y_i) \log(1 - p_i) \right].
\end{equation}

Another common example in real data analyses is Poisson GLMs, which can be applied to datasets with count dependent variables. In this case, the dependent variable $y_i$ follows a Poisson distribution with mean $\lambda_i$. The canonical link function is \( \log(\lambda_i) = \mathbf{x}_i^T \boldsymbol{\beta} \). Hence, the log-likelihood function is  \begin{equation}
    \ell(\boldsymbol{\beta}) = \sum_{i=1}^n \left[ y_i \log(\lambda_i) - \lambda_i - \log(y_i!) \right].
    \end{equation}

In addition to binary and count data, GLMs can be applied to various other types of data by specifying different link functions and distributions for the response variable. For instance, the identity link can model continuous data with a normal distribution and the reciprocal link can be used for gamma-distributed data. For detailed references on GLMs and their applications, see \cite{mccullagh2019generalized, dunn2018generalized}. Here, we focus on Bernoulli and Poisson GLMs.

\section{Variable Selection}\label{sec:3}

Variable selection is a crucial step in data analysis, and helps investigators better understand the underlying mechanisms behind the observed data. 
The goal of variable selection is to balance between model fit and model complexity. This perspective aligns with the principle of Occam’s razor~\citep{jefferys1992ockham}, which prefers to interpret the data by the most parsimonious model that explains the data well. This balance ensures that models are not too simple or unnecessarily complicated. 
To accomplish this, two primary aspects of variable selection need to be considered: assessment of model quality, and exploration of model space.

\subsection{Information criteria}
One way to evaluate the quality of a model is by using an information criterion. A good information criterion for variable selection should balance between model fit and model complexity. AIC and BIC are two common information criteria for variable selection, which can help address the trade-off between model fit and model complexity, and avoid overfitting or underfitting the data.

\subsubsection{Akaike information criterion}
 The AIC~\cite{akaike1974new,akaike1998information} is defined as:
\[
\text{AIC} = -2 \ell + 2k,
\]
where $\ell$ is the log-likelihood function of the model computed at the maximum likelihood estimator and $k$ is the number of parameters in the model.
A lower value of AIC indicates a better trade-off between model fit and model complexity.
The AIC can be applied to both LMs and GLMs.

\subsubsection{Bayesian information criterion}

The BIC~\cite{schwarz1978estimating}, also known as the Schwarz information criterion, is defined as:
\[
\text{BIC} = -2 \ell + k \log(n),
\]
where $\ell$ is the log-likelihood function of the model computed at the maximum likelihood estimator, $k$ is the number of parameters in the model, and $n$ is the number of observations (i.e., sample size).
Similarly to the AIC, a lower value of BIC indicates a better model. Both the AIC and BIC are penalized likelihood criteria, where the penalty term for AIC is $2k$, and for BIC is $k\log(n)$. When the sample size $n$ is large, BIC penalizes complexity more strictly and prefers smaller models than AIC.

\subsubsection{Other information criteria}

For models more complex than LMs and GLMs, there are other information criteria available. 
For example, the deviance information criterion (DIC, \cite{spiegelhalter2002bayesian}) balances model fit and model complexity for Bayesian hierarchical models. The Watanabe–Akaike information criterion (WAIC, \cite{watanabe2010asymptotic}) and widely applicable Bayesian information criterion (WBIC, \cite{watanabe2013widely}) are generalized versions of the AIC and BIC, respectively, for singular statistical models. In this paper, we focus on LMs and GLMs and, thus, do not consider the DIC, WAIC, and WBIC.

\subsection{Model search}

The model space is the set of  models that considers all possible combinations of regression variables in  
Equation (\ref{eq:linear-model}) for LMs or in the link Equation  (\ref{eq:glm-model}) for GLMs.
Thus, with $p$ candidate regression variables,  the model space is composed of $M=2^p$ models, which range in complexity from the null (i.e., empty) model  to the full model that includes all $p$ variables. 
There are many computational approaches to search the model space; next, we review some of the main approaches.

\subsubsection{Exhaustive search}
The only way to unequivocally determine the optimal model for a given information criterion is to compare all $M$ possible models in the model space. That is, to perform complete enumeration of all models to evaluate model quality. This is known as an exhaustive search of the model space.  
When $p$ is small (e.g., $p<16$ depending on sample size~\citep{williams2022bicoss}), an exhaustive search is feasible. If $p$ is large, exhaustive search is computationally expensive and potentially infeasible.  In this case, alternative approaches for searching larger model spaces should be considered.

\subsubsection{Greedy search} 
Greedy search methods make a series of locally optimal decisions at each iteration with the hope of finding a globally optimal model. Three common greedy search methods are forward selection,  backward elimination, and stepwise selection~\citep{hastie2009elements}.  
Forward selection starts with the null model (i.e., no regression variables) and adds one at a time, selecting the variable that improves the model the most based on an information criterion. Backward elimination starts with all $p$ regression variables and removes one at a time, removing the variable that reduces the information criterion the most. 
Stepwise selection combines forward selection and backward elimination by considering both the inclusion and exclusion of each regression variable at each iteration \citep{derksen1992backward}. When the  model space includes correlated regression  variables, the information criterion may have several local optima over the model space. In this case, greedy search methods may converge to a local optimum as opposed to a global optimum.

\subsubsection{Stochastic search}
Stochastic search uses a probabilistic approach to explore the model space. While also iterative, stochastic search algorithms are distinct from greedy approaches because they allow the selection of a model with a worse information criterion with some non-zero probability in a given iteration. This attribute of stochastic search allows the search to escape from local optima. 
When the model space is large and complex, stochastic search can solve a time-consuming problem in a feasible amount of time. One such example of a stochastic search approach is a genetic algorithm (GA; \cite{scrucca2013ga}), which emulates the process of natural evolution and is well suited to explore high dimensional discrete spaces.

\subsection{Regularization-based approaches}
 Regularization-based approaches, such as the LASSO, estimate the parameters of the model and search the model space in tandem. 
 To accomplish this, these approaches optimize an objective function that consists of a log likelihood term, as in Equation~\ref{eq:log-like-lm} for LMs and Equation~\ref{eq:log-like-glm} for GLMs, and a penalty term. 
 This penalty term shrinks the estimated coefficients of non-important regression variables to zero, thus resulting in variable selection.
 Because of the tandem nature of estimation and searching of the model space, regularization-based approaches are quite fast for dealing with problems with a large number of regression variables.

\subsubsection{LASSO} 
LASSO is a regularization based approach that uses an $L_1$ regularization penalty term~\cite{tibshirani1996regression}. For LMs, the  objective function to be minimized is:
\[
\sum_{i=1}^{n} \left( y_i - \mathbf{x}_i^\top \boldsymbol{\beta} \right)^2 + \lambda \sum_{j=1}^{p} |\beta_j|, 
\]
where
 $\sum_{i=1}^{n} \left( y_i - \mathbf{x}_i^\top \boldsymbol{\beta} \right)^2$ is the log-likelihood term,
    $\lambda \sum_{j=1}^{p} |\beta_j|$ is the $L_1$ regularization penalty term, and 
     $\lambda \geq 0$ is the regularization parameter that controls the strength of the penalty. A larger $\lambda$ shrinks more coefficients to zero, leading to a sparser model.

For  logistic regression, the objective function is:
\[
- \sum_{i=1}^n \left[ y_i \log(p_i) + (1 - y_i) \log(1 - p_i) \right] + \lambda \sum_{j=1}^{p} |\beta_j| ,
\]
 where the first term corresponds to the log-likelihood function, the second term is the penalty, and
$p_i = \exp(\mathbf{x}_i^\top \boldsymbol{\beta})/({1 + \exp(\mathbf{x}_i^\top \boldsymbol{\beta})})$ is the  probability of success for observation~$i$.
Similarly, for Poisson regression, the objective function is:
\[
- \sum_{i=1}^n \left[ y_i \log(\theta_i) - \theta_i \right]  + \lambda \sum_{j=1}^{p} |\beta_j| ,
\]
and $
\theta_i = \exp(\mathbf{x}_i^\top \boldsymbol{\beta})$ is the mean for observation~$i$.

Implementation of  LASSO requires selection of the regularization parameter $\lambda$ that multiplies the $L_1$ penalty. Selection of $\lambda$ can be performed using a grid-search of putative  regularization parameter values. Each one of these considered values corresponds to a potential model, which collectively compose the LASSO path.
Usually, the models in the LASSO path are compared based on predictive performance evaluated with cross-validation. 
Alternatively, they can also be compared with model selection criteria such as the AIC and BIC.

\subsubsection{Other methods}
There are several other regularization based approaches besides LASSO; for example, ridge regression~\citep{hoerl1970ridge} and  elastic net~\citep{zou2005regularization}.
See~\citep{friedrich2023regularization} for further discussion on other regularization based approaches.

\section{Simulations Studies}\label{sec:4}

To comprehensively evaluate the performance of each of the variable selection approaches, we consider six different combinations of  models (i.e., LMs, Bernoulli GLMs, and Poisson GLMs) and number of regression variables $p \in \{6,50\}$.
In each of these simulation studies, we consider a wide range of realistic sample sizes, effect sizes, and correlations among the regression variables. 
Because the conclusions are qualitatively consistent for all simulation studies, we discuss in this section only two simulation studies: (1) LMs with $p=6$, and (2) LMs with $p=50$. Results for the remaining four simulation studies are in {\bf Supplementary  Note 1}.

\subsection{Setup for Simulation Studies 1 and 2}
\paragraph{Data simulation}
We generate data from the following model:
\[
\boldsymbol{y} = \beta_0 \mathbf{1} + X \boldsymbol{\beta}+\boldsymbol{\epsilon}
\]
where $\boldsymbol{y}$ is a $n \times 1$ vector of response variable, $\beta_0 = 1$ is the intercept, $X$ is a $n \times p$ matrix for candidate variables,  $\boldsymbol{\beta}$ is a vector of corresponding regression coefficients, and $\boldsymbol{\epsilon}$ is an error vector distributed as  $N(\boldsymbol{0},\sigma^2\boldsymbol{I})$. 
In our simulation, a regression coefficient $\beta$ equals 1 if the variable is in the true model and 0 if the variable is not in the true model. With this, the relationship between the variance of the error $\sigma^2$ and the widely used Cohen's $f$ effect size is $f = 1/\sigma$ \citep{cohen2013statistical}.

To robustly explore the performance of these variable selection approaches, we consider $\sigma^2 \in \{6.25,16,100\}$  representing conventional large, medium, and small effect sizes~\citep{cohen2013statistical}, respectively. 
We consider sample sizes $n \in \{50, 100, 200,$ $400, 800, 1600, 3200, 6400\}$ to understand the interplay between effect size and sample size when performing variable selection. 
To evaluate how the different variable selection methods behave in the case of correlated regression variables, we consider correlations  $\rho \in \{0.00,0.10,0.25,0.50,0.75,0.90\}$. For each parameter setting of $\sigma^2$, $n$, and $\rho$ in a simulation study, we generate 100 datasets. 

\paragraph{Variable selection}
We implement the variable selection for the BIC and 
 the AIC differently for the cases when $p$ is small and $p$ is large.
When $p$ is equal to 6, there are 64 models in the model space, thus we perform variable selection using an exhaustive search of the model space. We refer to these two approaches as BIC and AIC.
When $p$ is equal to 50, there are about $10^{15}$ models, thus we perform a stochastic search GA. We refer to these approaches as GA\_BIC and GA\_AIC.  See \textbf{Supplementary Note 2} for details on the implementation of the GA.
For all simulation settings, we implement LASSO with the R package \texttt{glmnet}~\citep{glmnet2010}, where we use the default approach in \texttt{glmnet} to select the regularization parameter $\lambda$ that produces the lowest mean squared error (MSE) based on 10-fold cross-validation (CV). We refer to this approach as LASSO\_CV. In addition, we consider a sequence of regularization parameters $\lambda$ determined by the \texttt{glmnet} package, apply LASSO regression to fit the data for each $\lambda$, and subsequently evaluate the resulting models using the BIC or AIC to determine the optimal model. We refer to these approaches as LASSO\_BIC and LASSO\_AIC. Furthermore, we perform variable selection using a greedy stepwise approach combining both BIC and AIC for all settings. We refer to these approaches as Stepwise\_BIC and Stepwise\_AIC.

\paragraph{Performance evaluation}
We evaluate the variable selection approaches with three metrics. First, we consider the Correct Identification Rate (CIR) which is the number of simulated datasets for which the variable selection approach correctly selects the true model divided by the total number of simulated datasets. Second, we consider recall, which is the proportion of all true regression variables that are correctly selected. Third, we consider the False Discovery Rate (FDR), which is the number of selected regression variables that are not in the true model divided by the total number of selected regression variables. 
If the null model is selected for more than 50\% of the datasets, then the FDR is set to NA.
For each simulation setting, these three metrics are computed based on 100 simulated datasets.

\subsection{Simulation Results}
\paragraph{Simulation study 1: LMs, small number of regression variables}

\begin{figure}[htbp]
    \centering
   \includegraphics[width = 110mm]{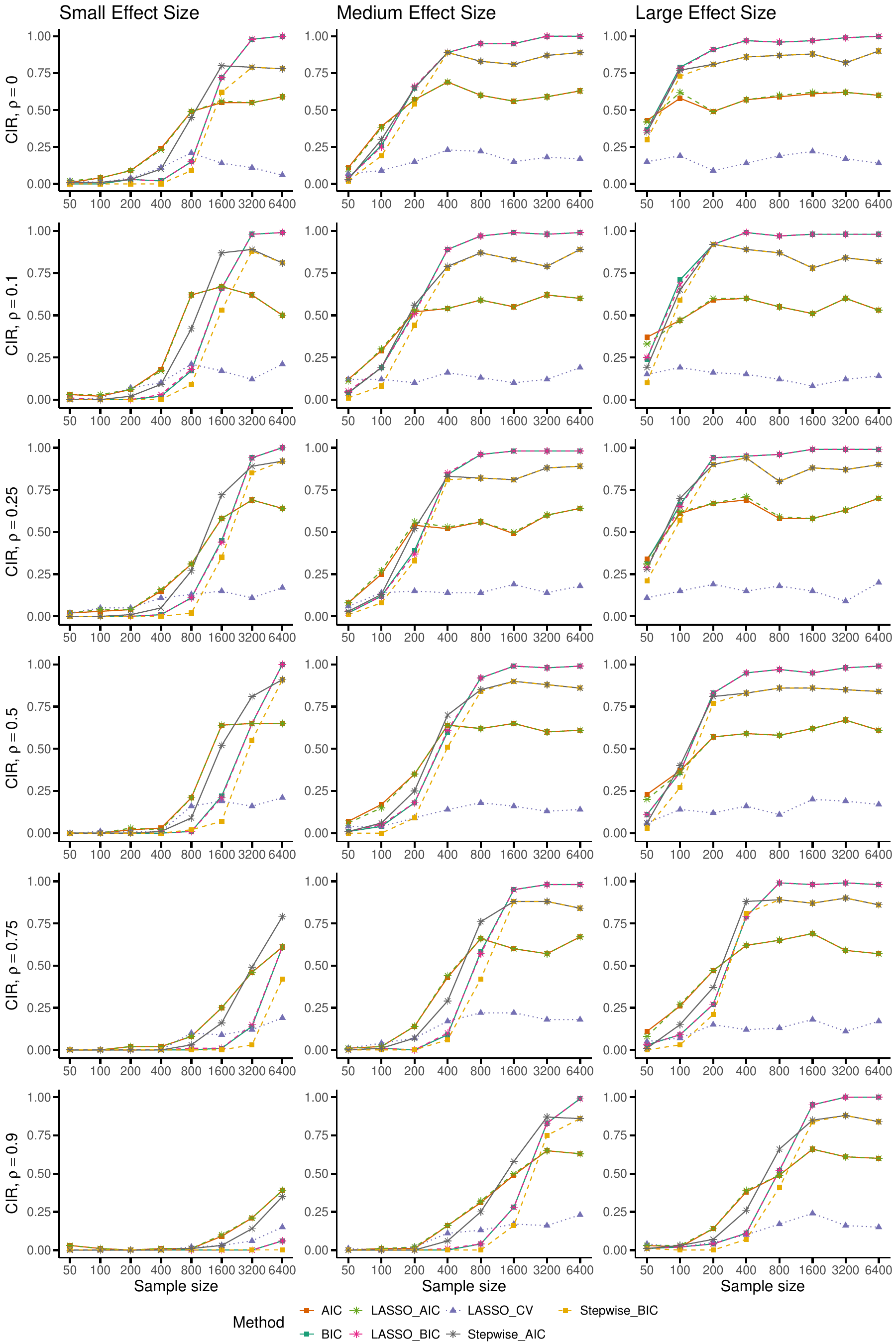}

    \caption{Simulation study 1 --- Comparison of CIR for BIC, AIC, LASSO\_BIC, LASSO\_AIC, LASSO\_CV, Stepwise\_BIC, and Stepwise\_AIC for continuous data with small number of regression variables and correlation $\rho \in \{0, 0.10, 0.25, 0.50, 0.75, 0.90\}$.}\label{fig:sim1_CIR}
\end{figure}

\begin{figure}[htbp]
    \centering
   \includegraphics[width = 110mm]{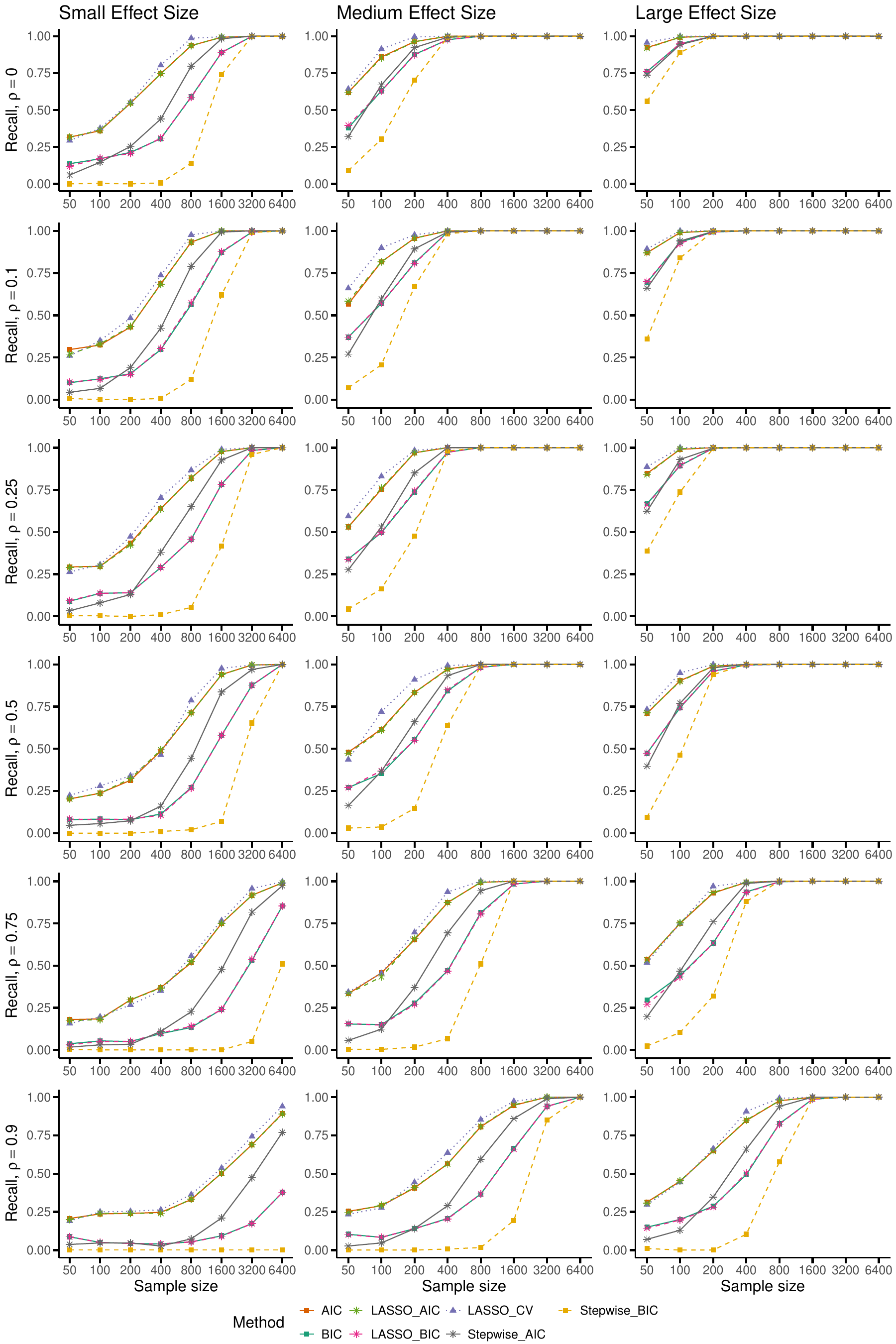}

    \caption{Simulation study 1 --- Comparison of recall for BIC, AIC, LASSO\_BIC, LASSO\_AIC, LASSO\_CV, Stepwise\_BIC, and Stepwise\_AIC for continuous data with small number of regression variables and correlation $\rho \in \{0, 0.10, 0.25, 0.50, 0.75, 0.90\}$.}\label{fig:sim1_recall}
\end{figure}

\begin{figure}[htbp]
    \centering
   \includegraphics[width = 110mm]{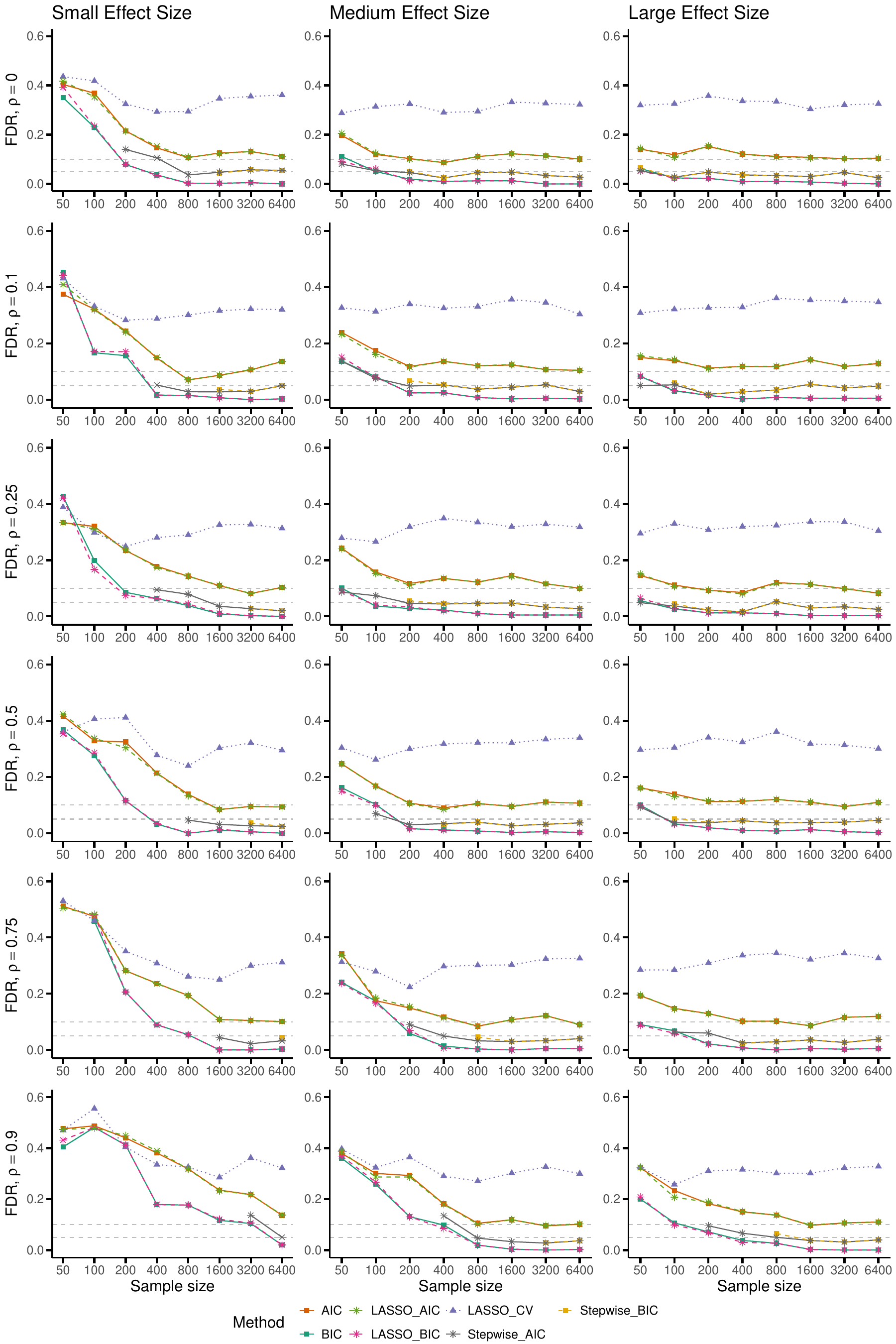}

    \caption{Simulation study 1 --- Comparison of FDR for BIC, AIC, LASSO\_BIC, LASSO\_AIC, LASSO\_CV, Stepwise\_BIC, and Stepwise\_AIC for continuous data with small number of regression variables and correlation $\rho \in \{0, 0.10, 0.25, 0.50, 0.75, 0.90\}$. Gray dashed lines are at 0.1 and 0.05. 
}\label{fig:sim1_FDR}
\end{figure}

In our first simulation study, we consider the case in which we have a LM with a small number of regression variables $p=6$. The model for simulating data is
\[
\mathbf{y} = \beta_0 \mathbf{1} + \beta_1 \mathbf{x_1} + \beta_2 \mathbf{x_2} + \beta_3 \mathbf{x_3} +\beta_4 \mathbf{x_4} +\beta_5 \mathbf{x_5} +\beta_6 \mathbf{x_6} +\boldsymbol{\epsilon},
\]
where $\mathbf{x_1},\mathbf{x_2},\mathbf{x_3}$ are the regression variables in the true model, and $\mathbf{x_4}, \mathbf{x_5}, \mathbf{x_6}$ are the regression variables that are not in the true model. We consider the setting when $\beta_1=\beta_2=\beta_3=1$, $\beta_4=\beta_5=\beta_6=0$, and the regression variables are simulated from the Gaussian distribution $N(\pmb{0}, \Sigma)$, where the diagonal elements of $\Sigma$ are equal to 1 and the off-diagonal elements are equal to $\rho$, with $\rho\in\{0.00,0.10,0.25,0.50,0.75,0.90\}$. 
We compare the performance of BIC, AIC, LASSO\_BIC, LASSO\_AIC, LASSO\_CV, Stepwise\_BIC and Stepwise\_AIC using CIR (\textbf{Figure~\ref{fig:sim1_CIR}}), recall (\textbf{Figure~\ref{fig:sim1_recall}}), and FDR (\textbf{Figure~\ref{fig:sim1_FDR}}) when considering large, medium and small effect sizes.

Based on the simulation study, we first observe that the CIR increases as sample size increases for all methods.
In the case of the BIC and LASSO\_BIC, CIR approaches 1.0 as the sample size increases. For the Stepwise\_BIC and Stepwise\_AIC, the CIR increases as sample size increases, but approaches an upper bound of CIR at approximately 0.9.
 AIC and LASSO\_AIC increase in CIR as sample size increases, but approaches an upper bound of CIR at approximately 0.69.
The LASSO\_CV reaches a maximum CIR of 0.23, regardless of setting and sample size.
Second, the rate to achieve maximum CIR depends on effect size.
 In the best case setting when $\rho=0$, sample sizes of 100-200, 400-800, and 1600-3200 may be required to achieve a close to maximal CIR for large, medium and small effect sizes, respectively.
Third, the rate to achieve maximum CIR depends on the correlation within the data.
 Specifically, as the correlation among regression variables increases, the rates to achieve maximum CIR are attenuated, with $\rho=0.9$ requiring the largest sample sizes.

Next, we consider the recall performance of these methods. Similar to CIR, recall increases as sample size increases for all methods. For medium and large effect sizes, all methods achieved a recall of 1 at sufficiently large sample sizes; larger sample sizes than explored here may be required in cases of small effect sizes. 
When comparing the methods,  LASSO\_CV exhibits greater recall at lower sample sizes  while Stepwise\_BIC consistently under-performs at recall compared to the other methods. 
 The rate to achieve maximum recall depends on effect size.
     Similar to CIR, in the best case setting when $\rho=0$,  sample sizes of 100-200, 200-400, and 1600-3200 may be required for large, medium and small effect sizes, respectively,  to achieve maximum recall.
 Lastly, as the correlation among regression variables increases, the sample sizes required to  achieve maximum recall also increase, with $\rho=0.9$ requiring the largest.
        
Finally, in our simulation studies we note that the FDR decreases as sample size increases for all methods, with exception of LASSO\_CV.
The BIC and LASSO\_BIC  methods approach an FDR of 0 as the sample size increases.
However, the FDR for Stepwise\_BIC and Stepwise\_AIC methods achieve a lower bound of 0.03; 
the AIC and LASSO\_AIC achieve a lower bound of 0.1;
 and the LASSO\_CV achieves a lower bound of 0.3. Note that for small sample sizes, Stepwise\_BIC and Stepwise\_AIC predominantly return the null model.
 The rate to achieve minimum FDR depends on effect size.
 In the best case setting when $\rho=0$, the  sample sizes to achieve minimum FDR for large, medium and small effect sizes are 50-100, 200-400, and 800-1600, respectively. These sample sizes are smaller than those needed to maximize CIR and recall.
Lastly, and similar to CIR and recall, the correlation among regression variables impacts the FDR, with  $\rho=0.9$ requiring the largest sample sizes
to minimize FDR.

\paragraph{Simulation study 2: LMs, high-dimension regression variables}
In our second simulation study, we seek to understand the variable selection characteristics of stochastic search when there is a large number of regression variables $p=50$. The model for simulating data is
\[
y_i = \beta_0 + \sum_{j=1\dots 50}\beta_j x_{ij} +\epsilon_i,
\]
where odd numbered regression variables $\mathbf{x_j} = (x_{1j},\dots,x_{nj}), j=1,3,5,\dots,49$ are in the true model, and even numbered regression variables $j=2,4,6,\dots,50$ are not in the true model. For each observation $i$, the regression variables are generated from an autoregressive model of order 1. That is, $x_{i1}$ follows $N(0, 1/(1-\rho^2))$ and, for $k=2,\dots,p$, $x_{ik} = \rho x_{i,k-1} + \eta_{ik}$, where $\eta_{ik}  \overset{\text{i.i.d.}}{\sim} N(0,1)$. Here, we set $\beta_j=1$ for $j=1,3,5,\dots,49$, and $\beta_j=0$ for $j=2,4,6,\dots,50$. Further, we consider settings where $\rho \in \{0.00,0.10,0.25,0.50,0.75,0.90\}$. Performance is evaluated for all the previously described methods from Simulation Study~1, except  GA\_BIC and GA\_AIC are used instead of exhaustive search BIC and AIC.

We observe many of the same conclusions as those in Simulation Study 1. 
As sample size increases, CIR tends to increase for GA\_BIC and LASSO\_BIC. In fact, under favorable conditions, the only approaches that near a CIR of 1 are GA\_BIC and LASSO\_BIC. All other methods reach a maximal CIR less than 1, with GA\_AIC  and LASSO\_CV consistently yielding a CIR close to 0.
Second, the rate at which maximal CIR is attained depends on effect size. Under the most favorable condition ($\rho=0$) for the GA\_BIC and LASSO\_BIC, sample sizes of roughly 3200, 6400, and more than 6400 are required to approach maximal CIR for large, medium, and small effect sizes, respectively. Third, this rate is also influenced by the correlation structure, with greater correlations requiring larger sample sizes to approach maximal CIR. Importantly,  the LASSO\_BIC is particularly susceptible to the adverse impact of high correlations among regressors, even at large sample sizes. In contrast, the CIR of the GA\_BIC quickly converges to 1 as the sample size increases.

Recall performance was consistent to that observed in Simulation Study 1.  Similar to CIR, recall increases with sample size across all methods with all methods achieving a recall of 1 at sufficiently large sample sizes for medium and large effect sizes and high correlation. 
Larger sample sizes than those explored here may be needed for small effect sizes. Under the considered conditions, LASSO\_CV exhibits higher recall at smaller sample sizes, while Stepwise\_BIC consistently under performs relative to the others. The rate at which maximum recall is reached depends on effect size: sample sizes of approximately 200, 400, and 1600–3200 may be required for large, medium, and small effect sizes, respectively, under the most favorable condition when $\rho=0$ for all methods. Lastly, as the correlation among regression variables increases, the sample sizes needed to achieve maximum recall also increase, with $\rho=0.9$ requiring the largest sample sizes.

Finally, in our simulation studies, we observe that the FDR decreases as sample size increases across all methods, except for LASSO\_CV. GA\_BIC  approaches an FDR of 0 as sample size grows. In contrast, Stepwise\_BIC and Stepwise\_AIC converge to a lower bound of approximately 0.04; GA\_AIC and LASSO\_AIC converge to an FDR of about 0.13; and LASSO\_CV remains at a FDR lower bound of 0.3. The rate at which minimum FDR is attained depends on effect size. Under the most favorable condition when $\rho=0$, sample sizes of roughly 200–400, 400–800, and 800–1600 may be required to reach minimum FDR for large, medium, and small effect sizes, respectively. Lastly, and similar to CIR and recall, the correlation among regression variables influences FDR, with $\rho=0.9$ requiring the largest sample sizes to minimize FDR.
Note that the FDR performance of LASSO\_BIC heavily depends on the correlation $\rho$, with the reductions in FDR requiring much larger sample sizes at higher correlations compared to GA\_BIC.

\begin{figure}[htbp]
    \centering
   \includegraphics[width = 110mm]{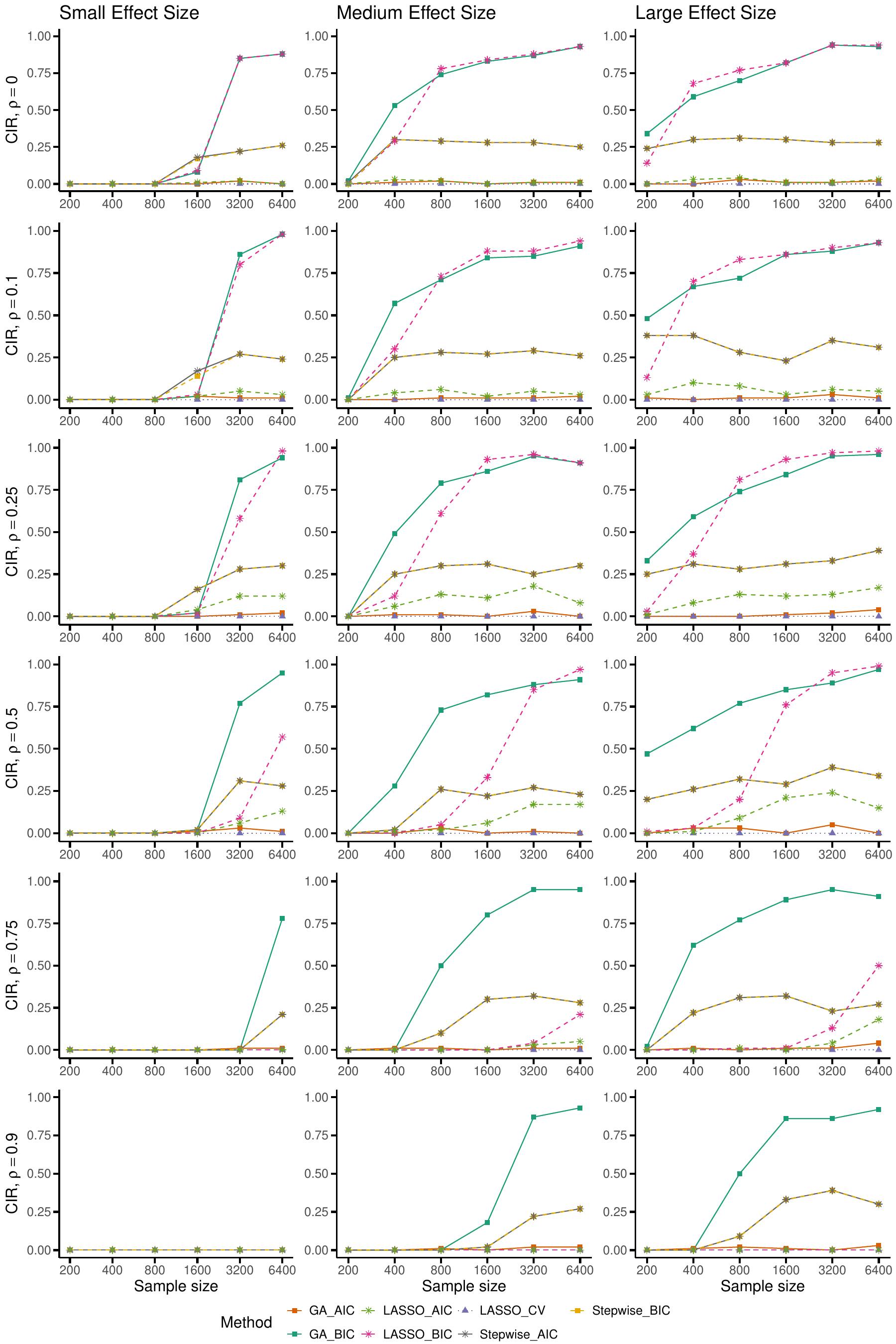}

    \caption{Simulation study 2 --- Comparison of CIR for GA\_BIC, GA\_AIC, LASSO\_BIC, LASSO\_AIC, LASSO\_CV, Stepwise\_BIC, and Stepwise\_AIC for continuous data with large number of regression variables and correlation $\rho \in \{0, 0.10, 0.25, 0.50, 0.75, 0.90\}$. Note the sample size starts at n=200.}\label{fig:sim2_CIR}
\end{figure}

\begin{figure}[htbp]
    \centering
   \includegraphics[width = 110mm]{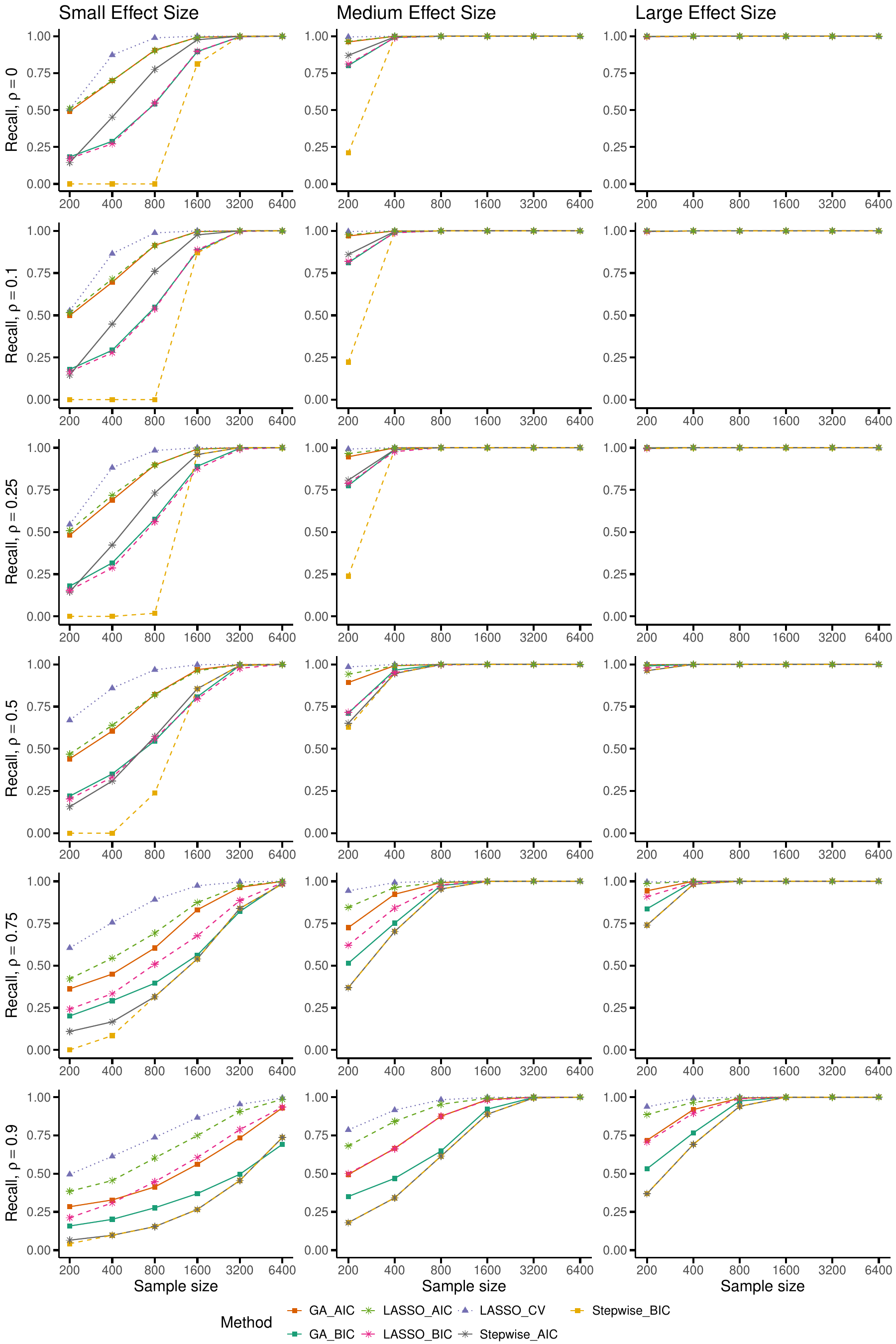}

    \caption{Simulation study 2 --- Comparison of recall for GA\_BIC, GA\_AIC, LASSO\_BIC, LASSO\_AIC, LASSO\_CV, Stepwise\_BIC, and Stepwise\_AIC for continuous data with large number of regression variables and correlation $\rho \in \{0, 0.10, 0.25, 0.50, 0.75, 0.90\}$. Note the sample size starts at n=200.}\label{fig:sim2_recall}
\end{figure}

\begin{figure}[htbp]
    \centering
   \includegraphics[width = 110mm]{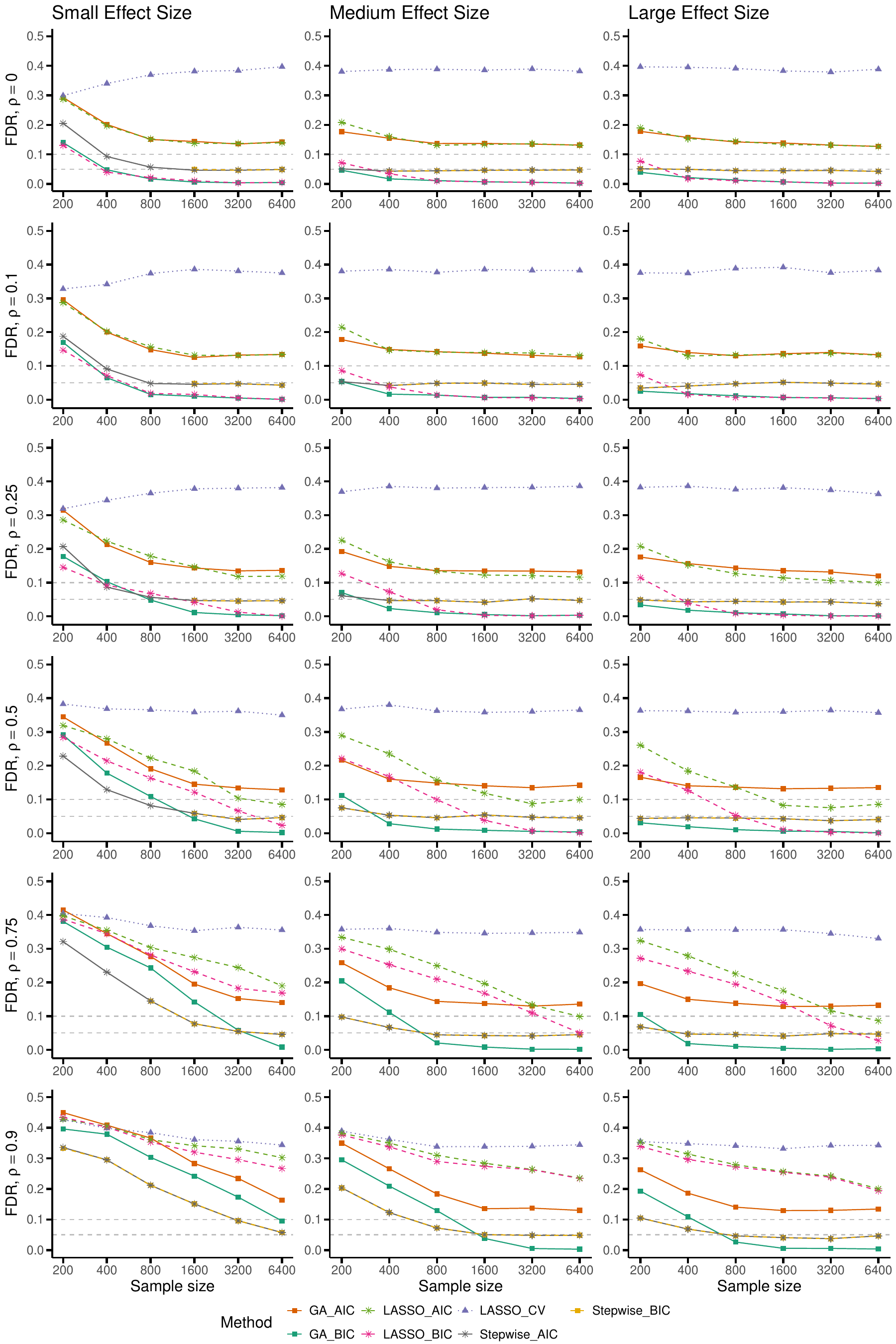}

    \caption{Simulation study 2 --- Comparison of FDR for GA\_BIC, GA\_AIC, LASSO\_BIC, LASSO\_AIC, LASSO\_CV, Stepwise\_BIC, and Stepwise\_AIC for continuous data with large number of regression variables and correlation $\rho \in \{0, 0.10, 0.25, 0.50, 0.75, 0.90\}$. Note the sample size starts at n=200. Gray dashed lines are at 0.1 and 0.05. }\label{fig:sim2_FDR}
\end{figure}

\section{\texttt{modelselection} package in R} \label{sec:package}
To help facilitate the use of the BIC for variable selection, we have developed the R package \texttt{modelselection}. This package implements complete enumeration of the model space for small $p$, and a GA stochastic search of the model space for large $p$. \texttt{modelselection} works for both LMs and GLMs, and uses the same formula syntax of the \texttt{lm} and \texttt{glm} functions in base \texttt{R}. This package can be accessed online at \href{https://github.com/xss55/modelselection}{GitHub repository (https://github.com/xss55/modelselection)} and will be made available on CRAN upon publication. In addition, we provide the code necessary to reproduce our simulation studies as {\color{blue} Supplementary Material.}

\section{Discussion} \label{sec:5}

A variable selection method consists of model evaluation using an information criterion and a search of the model space. In this paper, we considered the BIC and AIC for evaluating models, and  exhaustive, greedy, LASSO path, and  GA approaches for searching the model space; we also considered LASSO using cross validation.  
We comprehensively compared and evaluated these nine variable selection methods on simulated datasets using the performance metrics of CIR, recall, and FDR.
We explored relevant dimensions in our simulation studies including the sample size, effect size, and correlation among predictors. 
We found that most BIC-based methods  achieved a maximum CIR of 1, while other methods resulted in reduced CIR.
All methods were able to recall the correct variables in the model, but at different rates based on sample size, effect size, and correlation. 
Lastly, only BIC-based approaches were able to achieve an FDR of 0.

An important property for variable selection methods is to consistently select the correct/true model, a property commonly referred to as model selection consistency~\citep{vrieze2012model,shao1997asymptotic}. 
This property describes variable selection methods for which the probability of correctly selecting the true model converges to one as the sample size increases. Here, we use the CIR to represent the property of model selection consistency.
In our simulation studies, the BIC and GA\_BIC approaches  achieved the highest CIR of all methods considered as the sample size increased.
This excellent behavior of these two approaches is a direct consequence of the BIC being model selection consistent~\citep{vrieze2012model}.

The lack of replicability of research findings remains an outstanding challenge. 
Replicability in variable selection means the ability to select the same variables based on different sample datasets from the same population. 
In our simulation studies, we use the FDR to represent replicability. 
While all approaches resulted in high recall, both the exhaustive search BIC and the GA\_BIC approaches had the lowest FDR. This observation is further supported by the penalization term in the BIC that penalizes complexity more strictly and prefers smaller models than the AIC.
Collectively, reducing the chance of identifying a variable as important that is truly not important may result in increasing the replicability of research findings from one study to another, and aid in preventing spurious findings.

For many scientific questions, such as in genetics \citep{williams2022bicoss,xu2023bg2,xu2025genome} and psychology \citep{craft2023long, tegge2025treatment,downey2025s}, understanding the underlying mechanism is the goal. Our simulation studies have led us to make the following recommendations to researchers. 
For model spaces with a small number of regression variables, an exhaustive search of the model space is feasible. Moreover, model evaluation with the BIC using either the exhaustive search or the LASSO path elicited maximal CIR. Stepwise-based search approaches, and AIC-based model evaluation were suboptimal. These findings hold regardless of explored correlation structures. Note that for all sample sizes, the BIC using either the exhaustive search or the LASSO path have the lowest FDR, thereby making conservative decisions regarding the variables to be considered as important. 

For model spaces with a large number of regression variables, an exhaustive search of the model space is no longer feasible. The GA\_BIC and the LASSO\_BIC permit search of this complex model space and perform with highest CIR. However,  CIR for the LASSO\_BIC is severely impacted as the correlation among regression variables increases, due to high FDR. 
All other methods failed to achieve a high CIR  or a low FDR, regardless of effect size, sample size, or correlation.
For all simulation studies, the traditional implementation of LASSO with cross validation performs poorly due to its high FDR.

There are many possible avenues for future research. In the future, we plan to extend our exploration of variable selection  to linear mixed models (LMM) and generalized linear mixed models (GLMM). Research exists regarding variable selection for LMMs and GLMMs \citep{xu2023bayesian}. We will explore these approaches to characterize the interplay between model evaluation and model search. Furthermore, we plan to extend our study of variable selection methods to survival analysis and ultra-high dimensional problems such as genome-wide association studies \citep{williams2022bicoss,xu2023bg2}.

\section{CREDIT}

SX:  Conceptualization, Software, Validation, Formal analysis, Investigation, Data Curation, Writing - Original Draft, Writing - Review \& Editing, Visualization

MARF: Conceptualization, Formal analysis, Investigation, Writing - Original Draft, Writing - Review \& Editing

ANT: Conceptualization, Formal analysis, Investigation, Supervision, Writing - Original Draft, Writing - Review \& Editing

\bibliography{ref}

\end{document}